\documentclass[prd,preprint,superscriptaddress,amsmath,amssymb,nofootinbib]{revtex4}
\usepackage[compat=1.1.0]{tikz-feynman}
\usepackage{graphicx}
\usepackage{dcolumn}
\usepackage{bm}
\usepackage{amssymb}
\usepackage{amsmath}
\usepackage{epsfig}    
\usepackage{color}
\usepackage{slashed}
\usepackage{hhline}
\usepackage{bbm}

\def\be{\begin{equation}}
\def\ee{\end{equation}}
\newcommand{\bea}{\begin{eqnarray}}
\newcommand{\eea}{\end{eqnarray}}
\newcommand{\nn}{\nonumber}

\newcommand{\ys}[1]{{\textcolor{orange}{[YS: {#1}]}}}

\begin{document}

\title{A novel realization of linear seesaw model in a non-invertible selection rule with the assistance of $\mathbb Z_3$ symmetry }

\author{ Hiroshi Okada}
\email{hiroshi3okada@htu.edu.cn}
\affiliation{Department of Physics, Henan Normal University, Xinxiang 453007, China}

\author{Yutaro Shoji}
\email{yutaro.shoji@ijs.si}
\affiliation{J$\hat oo$zef Stefan Institute, Jamova 39, 1000 Ljubljana, Slovenia}

\date{\today}

\begin{abstract}
We propose a novel realization of linear seesaw model in a non-invertible selection rule with the assistance of $\mathbb Z_3$ symmetry. In our framework, Dirac mass matrices are generated at one-loop level, dynamically breaking the non-invertible symmetry while the symmetry is invariant under the tree-level.
In addition to the active neutrino masses, the model exhibits rich and testable phenomenology such as non-unitarity bound, lepton flavor violations, lepton anomalous magnetic moment, and dark matter candidate.
After describing our model, we carry out numerical analysis and show some results for our physical parameters.  
\end{abstract}
\maketitle

\section{Introduction}
The existence of tiny but nonzero active neutrino masses provides a clear hint of physics beyond the Standard Model (BSM),
and vast amount of literature has been proposed to explain the smallness of the neutrino masses. 
Along these lines,
the Linear SeeSaw (LSS)~\cite{Wyler:1982dd, Akhmedov:1995ip, Akhmedov:1995vm} and the Inverse SeeSaw (ISS)~\cite{Mohapatra:1986bd, Wyler:1982dd} are well-known verifiable scenarios generating tiny neutrino masses within TeV scale, predicting tiny Majorana mass matrices for sterile neutral fermions.
Their minuscule masses are supported by conservation of lepton number as the violation of lepton number induces the tiny active neutrino masses.
This type of mechanisms explaining small parameters is referred to as `technically naturalness' in the 't Hooft sense.
The model includes left-handed and right-handed neutral fermions; $N_R$ and $N_L$.
After spontaneous electroweak symmetry breaking, one generally writes the following mass terms:
\begin{align}
m_{D} {\overline N_R} \nu_L +m'_{D} {\overline N_L^C} \nu_L +
M_N {\overline N_R}N_L + \mu_R {\overline N_R} N_R^C +\mu_L {\overline N_L^C} N_L +{\rm c.c.}\ , \label{eq:iss}
\end{align}
where $m_{D}\equiv y_d v/\sqrt2$ and $m'_D \equiv y'_d v/\sqrt2$.
Here, $v$ is vacuum expectation value (VEV)  of the SM Higgs boson $H$ denoted by $\langle H\rangle\equiv [0,v/\sqrt2]^T$.
The mass matrix for the neutral fermions in the basis $[\nu_L^C,N_R, N_L^C]^T$ is given by
\begin{align}
\begin{pmatrix}
 0 & m_D & m'_D  \\
 m_D^T & \mu_R & M_N^T \\
 m'^T_D & M_N & \mu_L 
  \end{pmatrix}. \label{neut-mass}
\end{align}
If we impose the mass hierarchies as
\begin{align}
 \mu_L,\ \mu_R \ll
   m_D,\ m'_D <  M_N, \label{eq:mass-order}
\end{align}
one finds the following active neutrino mass matrix:
\begin{align}
 m_\nu&= m'_{D} (M_N^T)^{-1}  m^T_D +m_{D} M_N^{-1}  m'^T_D \label{eq:Linear}\\
&\hspace{3ex} -m_{D} M_N^{-1} \mu_L  (M_N^T)^{-1}  m^T_D - m'_{D} (M_N^T)^{-1} \mu_R M_N^{-1}  m'^T_D.
\label{eq:Inverse}
\end{align}
If the first two terms dominate, 
the model is called LSS; whereas, if the last two terms dominate,
the model is called ISS.
Given that the active neutrinos mix with the moderately heavy neutral fermions, non-unitarity constraints become relevant: one has to take into account several experimental results such as the effective Weinberg angle, SM $W$ boson mass, several ratios of $Z$ boson fermionic decays, invisible decay of $Z$, electroweak universality, measurements of the quark mixing matrix, and lepton flavor violations (LFVs).
The constraints are summarized in terms of $\epsilon\equiv m'^*_D (M^\dag_N)^{-1} M^{-1}_N m'^T_D+m^*_D (M^*_N)^{-1} (M^T_N)^{-1} m^T_D$~\cite{Agostinho:2017wfs}:
\begin{align}
|\epsilon|\lesssim 
\left[\begin{array}{ccc}
4.08\times10^{-5}  & 1.65\times10^{-5}  & 5.19\times10^{-5}  \\ 
1.65\times10^{-5} & 3.85\times10^{-5} & 5.04\times10^{-5} \\ 
5.19\times10^{-5} & 5.04\times10^{-5} &1.12\times10^{-4} \\ 
\end{array}\right],
\end{align}
where $|\epsilon|$ indicates the absolute values of the matrix elements.
%
%
%
These constraints set the scale of $m_D$ and $m'_D$ to be much smaller than that of $M_N$:
\begin{align}
   ||m_D||,\ ||m'_D|| \ll  ||M_N||. \label{eq:neutmass-order_exp}
\end{align}
However, from a theoretical perspective, there is no satisfactory explanation as long as one remains within the simplest model.

In this paper, we propose a theoretical explanation for Eq.~(\ref{eq:neutmass-order_exp}) by introducing a non-invertible symmetry known as the $\mathbb Z_3$ gauging Tambara-Yamagami fusion rule\footnote{See Appendix in details. Along this line of idea, several ideas have recently been applied to phenomenology~\cite{Jiang:2025psz, Kobayashi:2025rpx, Jangid:2025krp, Okada:2025kfm, Choi:2022jqy,Cordova:2022fhg, Cordova:2022ieu, Nomura:2025tvz,Cordova:2024ypu,Kobayashi:2024cvp,Kobayashi:2024yqq,Kobayashi:2025znw,Suzuki:2025oov,Liang:2025dkm,Kobayashi:2025ldi,Kobayashi:2025lar,Kobayashi:2025cwx,Nomura:2025sod,Dong:2025jra,Nomura:2025yoa,Chen:2025awz, Kobayashi:2025thd, Suzuki:2025bxg}. }, assisted by a discrete Abelian symmetry $\mathbb Z_3$ and new particles BSM. In our framework, the mass terms $m_D$ and $m'_D$ arise at the one-loop level as leading contributions\footnote{Similar ideas have been appeared in refs.~\cite{Nomura:2018ktz, Nomura:2024jxc, Nomura:2021adf, Okada:2012np, Binh:2024lez, Bonilla:2023egs, Bonilla:2023wok, Das:2017ski}.}, whereas $M_N$ is allowed at the tree level.
This setup leads to rich and verifiable phenomenology such as LFVs, the lepton anomalous magnetic moment of the lepton (lepton $g-2$), and a dark matter (DM) candidate.
We perform numerical analyses incorporating all relevant phenomenological constraints, and present the allowed parameter space of our model for both normal hierarchy (NH) and inverted hierarchy (IH) of active neutrino masses.

This paper is organized as follows. In Section~II, we present our setup, and formulate the neutral fermion mass matrices, LFVs and lepton $g-2$, and dark matter candidate and its cross section to explain the observed the relic density.
In Section III, we perform numerical analysis to search for allowed parameter region and illustrate the resulting physical tendencies. 
Lastly, we devote Section~IV to the summary and conclusion.
We also provide a review of $\mathbb Z_3$ gauging TY fusion rule in Appendix A.

\begin{table}[!ht]
\begin{center}
\begin{tabular}{|c||c|c|c|c|c|c||c|c|c|}\hline\hline
& \multicolumn{6}{c||}{Fermions} & \multicolumn{3}{c|}{Bosons} \\ \hline
Fields & ~$L_L$~ & ~$\ell_R$~ & ~$N_R$~ & ~$N_L$~ & ~$X_R$~& ~$X'_R$~  & ~~$H$~ & ~$\eta$~ & ~$S_0$~  \\ \hline
$SU(2)_L$ & $\mathbbm{2}$ & $\mathbbm{1}$ & $\mathbbm{1}$  & $\mathbbm{1}$   & $\mathbbm{1}$ & $\mathbbm{1}$ & $\mathbbm{2}$
& $\mathbbm{2}$ & $\mathbbm{1}$ \\ \hline
$U(1)_Y$ & $-\frac{1}{2}$ & $-1$ & $0$ & $0$& $0$  & $ 0$ & $ \frac{1}{2}$ & $\frac{1}{2}$ & $0$ \\ \hline
${\rm TY}$ & $\mathbbm{1}$ & $\mathbbm{1}$& $c$  & $c$  & $n$ & $n$ & $\mathbbm{1}$ & $n$& $n$ \\ \hline
$Z_{3}$ & $1$ & $1$ & $1$ & $1$  & $\omega$& $\omega^2$  & $1$& $\omega$& $\omega$ \\ \hline 
\end{tabular}
\caption{Field contents and their charge assignments under the $SU(2)_L\otimes U(1)_Y\otimes {\rm TY}\otimes \mathbb Z_3$.} 
\label{tab:fieldcontent}
\end{center}
\end{table}

\section{Model}
\label{sec:II}
\subsection{Setup}
Here, we review our model.
In addition to the SM fields, we introduce neutral fermions $N_R$, $N_L$, $X_R$ and $X'_R$, an inert doublet boson $\eta(\equiv [\eta^+,\eta^0]^T)$ and a singlet inert boson $S_0$, where all neutral fermions are supposed to have three families for simplicity.
Here, $N_L$ and $N_R$ are relevant to our linear seesaw model, whereas $X_R,\ X'_R$, $\eta,\ S_0$ play a role in generating Dirac mass matrice $m_D$ and $m'_D$ at one-loop level after dynamically violating the TY rule. 
The SM Higgs is denoted by $H\equiv[w^+,(v+h+i z)/\sqrt2]^T$, where $w^+$ and $z$ are absorbed by the SM gauge bosons $W^+$ and $Z_0$, respectively.
%
We impose TY and $\mathbb Z_3$ charges to these fields in order to realize the desired model structure.
Their field contents and charge assignments are summarized in Table \ref{tab:fieldcontent}.
Under these symmetries, the Lagrangian in the lepton sector has the following renormalizable terms: 
\begin{align}
&y^\ell_{ii} \overline{L_{L_i}} H \ell_{R_i} + M_{N_{ii}} \overline{N_{L_i}} N_{R_i} + y^\eta_{i\alpha} \overline{L_{L_i}} \tilde \eta X_{R_\alpha}\label{eq:1}\\
&+ y'^R_{a\alpha} \overline{N^C_{R_a}} X'_{R_\alpha} S_0 
+ y^R_{a\alpha} \overline{N^C_{R_a}} X_{R_\alpha} S_0^*
+ y'^L_{a\alpha} \overline{N_{L_a}} X'_{R_\alpha} S_0 
+ y^L_{a\alpha} \overline{N_{L_a}} X_{R_\alpha} S_0^*
+ M_{X_{\alpha\alpha}} \overline{X^C_{R_\alpha}} X'_{R_\alpha}+ {\rm h.c.},
\label{eq:2} 
\end{align}
where $\tilde\eta\equiv i\sigma_2\eta^*$ with $\sigma_2$ being the second Pauli matrix.
Here, we take $y^\ell,\ M_N,\ M_X$ to be diagonal matrices without loss of generality. There is one important term in the scalar potential in order to induce small Dirac mass matrices:
\begin{equation}
\mu [(\eta^\dag H)S_0+{\rm c.c.} ]. 
 \end{equation}

The TY fusion rule plays an essential role in structuring our linear seesaw model. Below, we list up the terms forbidden by the TY rule that would spoil our mass generation mechanism:
\begin{align}
&\overline{L_L}\tilde HN_R,\quad \overline{L_L}\tilde HN^C_L,\quad \overline{L_L}\tilde H X_R,\quad \overline{L_L}\tilde H X'_R,\quad
 \overline{L_L}\tilde\eta N_R,\quad \overline{L_L}\tilde\eta N^C_L,\\ 
 &\overline{N^C_R}N_R,\quad \overline{N^C_R}N_R S_0^{(*)},\quad \overline{N^C_L}N_L,\quad \overline{N_L}N_L S_0^{(*)},\quad
 \overline{X^C_R} X_R S_0^{(*)},\quad \overline{X'^C_R} X'_R S_0^{(*)},  \\
& \overline{X^C_R} X'_R S_0^{(*)},\quad S_0,\quad S_0^3,\quad (\eta^\dag \eta)(\eta^\dag H).
\end{align}
However, the TY fusion rule still allows the following terms:
\begin{align}
S^2_0,\quad (\eta^\dag H)^2, \quad (H^\dag\eta)S_0, \quad  \overline{X^C_R} X_R ,\quad  \overline{X'^C_R} X'_R , \quad \overline{L_L}\tilde\eta X'_R.
\end{align}
With these terms, our model becomes a radiative seesaw model instead of the LSS model.
This is why we have introduced the additional $\mathbb Z_3$ symmetry, which forbids these terms.

In addition, we set $y'^{L,R}=0$ in the following analysis. Phenomenologically, these terms are harmless since they only generate $\mu_{L,R}$ in Eq.~(\ref{neut-mass}) at the one-loop level, giving higher order corrections to the active neutrino mass matrix.
Theoretically, however, these one-loop diagrams are divergent and necessitate the counter terms. Since such counter terms are forbidden by the TY rule, our model would be spoiled.
This problem is resolved by following argument. If $y'^{L,R}=0$ and $M_X=0$, $X'_L$ and $X'_R$ decouple from the theory and an additional symmetry is recovered. Since $M_X$ is a dimensionful parameter, it can be generated by a soft breaking of the symmetry in an extended model, while maintaining $y'^{L,R}=0$.
This structure also guarantees that, once we set $y'^{L,R}=0$ in our model, they are not generated by quantum corrections.

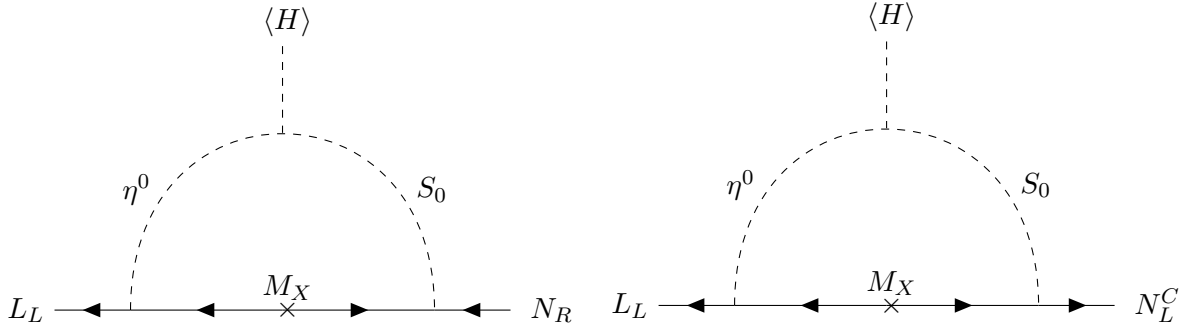
\begin{figure}
    \centering
\begin{tikzpicture}
  \begin{feynman}
    \vertex[label=left:\(\ L_L \)] (a) at (0,0);       
    \vertex[label=right:\(\ N_R \)] (b) at (6,0);       
    \vertex (c) at (1,0);       
    \vertex[crossed dot, label=above:\(\ M_X\)] (m) at (3,0);
    \vertex[label=below:\(\ \times\), yshift=7.5pt] at (3,0);       
    \vertex (d) at (5,0);       
    \vertex (e) at (3,2.35);       
    \vertex[label=above:\(\ \langle H\rangle\)] (f) at (3,3.5);     %
    \vertex[label=left:\(\ \eta^0\)] (h) at (1.4,1.6);       
    \vertex[label=right:\(\ S_0\)] (i) at (4.5,1.6);       
    \diagram* {
      (c) -- [fermion] (a),    
      (m) -- [fermion] (c),    
      (m) -- [fermion] (d),    
      (b) -- [fermion] (d),
      (c) -- [scalar, half left, looseness=2] (d),  
      (f) -- [scalar] (e),                         
    };
  \end{feynman}
\end{tikzpicture}
\begin{tikzpicture}
  \begin{feynman}
    \vertex[label=left:\(\ L_L \)] (a) at (0,0);       
    \vertex[label=right:\(\ N_L^C \)] (b) at (6,0);       
    \vertex (c) at (1,0);       
    \vertex[crossed dot, label=above:\(\ M_X\)] (m) at (3,0);
    \vertex[label=below:\(\ \times\), yshift=7.5pt] at (3,0);       
    \vertex (d) at (5,0);       
    \vertex (e) at (3,2.35);       
    \vertex[label=above:\(\ \langle H\rangle\)] (f) at (3,3.5);     %
    \vertex[label=left:\(\ \eta^0\)] (h) at (1.4,1.6);       
    \vertex[label=right:\(\ S_0\)] (i) at (4.5,1.6);       
    \diagram* {
      (c) -- [fermion] (a),    
      (m) -- [fermion] (c),    
      (m) -- [fermion] (d),    
      (d) -- [fermion] (b),
      (c) -- [scalar, half left, looseness=2] (d),  
      (f) -- [scalar] (e),                         
    };
  \end{feynman}
\end{tikzpicture}
    \caption{One-loop diagrams for the Dirac mass matrices, $m_D$ (left) and $m'_D$ (right).}
    \label{fig:mD}
\end{figure}

\subsection{The Dirac mass matrices and active neutrino mass matrix}
Here, we discuss the Dirac mass matrices $m_D$ and $m'_D$, which are generated at the one-loop level by the diagrams in Fig~\ref{fig:mD}.
The resulting contributions are given by
\begin{align}
&(m_D)_{ij} = \sum_\alpha \frac{\mu^*vM_{X_\alpha} }{16\sqrt2\pi^2} y^\eta_{i\alpha} F(M_{X_\alpha},m_{\eta^0},m_{S_0}) y^R_{j\alpha},  \label{eq:mD}  \\
&(m'_D)_{ij} = \sum_\alpha \frac{\mu^*vM_{X_\alpha} }{16\sqrt2\pi^2} y^\eta_{i\alpha}F(M_{X_\alpha},m_{\eta^0},m_{S_0})y^L_{j\alpha},  \label{eq:mD'}  \\
& F(m_1,m_2,m_3)\approx \frac{1}{m^2_3-m^2_2}
\left[
\frac{m^2_3}{m^2_3-m^2_1}\ln\left(\frac{m^2_3}{m^2_1}\right)
-
\frac{m^2_2}{m^2_2-m^2_1}\ln\left(\frac{m^2_2}{m^2_1}\right)
\right].
\end{align}

Our active neutrino mass matrix is then given at two-loop level through linear seesaw terms in Eq.~(\ref{eq:Linear}):
\begin{align}
m_\nu =  m'_{D} M_N^{-1}  m^T_D +m_{D} M_N^{-1}  m'^T_D,
\end{align}
%
%
%
which is diagonalized by a unitary matrix $U_\nu$ as
$
(D_1, D_2, D_3)=U^\dag_\nu m_\nu U^*_\nu$.

{Let us impose the following condition for simplicity:
\begin{align}
y\equiv y^L=y^R,
\end{align}
which results in $m_D=m'_D$.
Then, we can apply the Casas-Ibarra parametrization~\cite{Casas:2001sr} and use the following expression:
\begin{align}
m_D=\frac{1}{\sqrt2} U D^{1/2}_\nu {\cal O} M_N^{1/2},
\end{align}
where ${\cal O}$ is three by three orthogonal mixing matrix with complex values; ${\cal O}^T{\cal O}={\cal O}{\cal O}^T=1$.
Since we take the diagonal basis for the charged-lepton mass matrix, we have
$U\equiv U_{\rm PMNS}= U_\nu$, where $U_{\rm PMNS}$ is the Pontecorvo–Maki–Nakagawa–Sakata (PMNS) matrix~\cite{Maki:1962mu} with the standard parametrization.

It enables us to solve $y^\eta$ as
\begin{align}
y^\eta_{i\alpha}=\frac{16\pi^2}{\mu^*v}\frac{(U_\nu D^{1/2}_\nu {\cal O} M^{1/2}_N {(y^T)}^{-1})_{i\alpha}}{M_{X_\alpha} F(M_{X_\alpha},m_{\eta^0},m_{S_0})},\label{eq:ci}
\end{align}
where we require it to satisfy the perturbative limit $|y^\eta|\lesssim \sqrt{4\pi}$.
}

The sum of the neutrino masses $\sum_i D_i$ has an upper bound of 120 meV~\cite{Vagnozzi:2017ovm, Planck:2018vyg} assuming
the minimal cosmological model.
Meanwhile, the recent result for $\Lambda$CDM is $\sum D_{i}\le$ 72 meV~\cite{DESI:2024mwx} obtained from the combined result of DESI and CMB. 

 The effective neutrino mass for the neutrinos double beta decay, which is denoted by $m_{ee}$, is given by
   \begin{align}
m_{ee}
=
\left|
D_1 c^2_{12} c^2_{13} + D_2 s^2_{12}c^2_{13} e^{i\alpha} + D_3 s^2_{13} e^{i(\beta-2\delta_{\rm CP})}
\right| .
\end{align}
It has  upper bounds $(28–122 )$ meV at 90\% confidence level from
a current KamLAND-Zen data~\cite{KamLAND-Zen:2024eml}. 
On the other hand, the effective electron antineutrino mass is defined by
\begin{align}
m_{\nu e}\equiv \sum_i D_i^2|U_{\nu_{1i}}|^2
=
D_1^2 c^2_{12} c^2_{13} + D_2^2 s^2_{12}c^2_{13}  + D_3^2 s^2_{13}.
\end{align}
It is bounded from the global analysis of oscillation data together with the KATRIN experiment at 95\% CL~\cite{Esteban:2024eli}:
\begin{align}
& {\rm NH}:\ 0.85{\rm meV}\le m_{\nu e}\le 400 {\rm meV},\\
& {\rm IH}:\ 48{\rm meV}\le m_{\nu e}\le 400 {\rm meV}.
\end{align}

\subsection{LFVs and lepton $g-2$}
\label{sec:leptonpheno}
The coupling $y^\eta$ between the charged leptons and new particles induces contributions to LFVs and lepton $g-2$. 
The branching ratios of $\ell_i \to \ell_j \gamma$ are calculated as
\begin{align}
&\frac{{\rm BR}(\ell_i \to \ell_j \gamma)}{{\rm BR}(\ell_i \to \nu_i \bar\nu_j \ell_j)}
\approx\frac{3\alpha_{\rm em}}{16\pi {\rm G_F}^2}
\left|\sum_\alpha
y^\eta_{j\alpha} y^{\eta*}_{i\alpha} G(M_{X_\alpha},m_{\eta^+})
\right|^2,\\
&G(m_1,m_2)
\simeq 
\frac{2 m^6_1+3m_1^4m_2^2-6m_1^2m_2^4+m_2^6+6m_1^4m_2^2\ln\left(\frac{m_2^2}{m_1^2}\right)}{12(m_1^2-m_2^2)^4},
\end{align}
where $\alpha_{\rm em}\approx1/137$ is the fine-structure constant,
${\rm BR}(\ell_i \to \nu_i \bar\nu_j \ell_j) \approx(1,0.1784, 0.1736)$ for ($(i,j)=(\mu,e),(\tau,e),(\tau,\mu)$), and 
${\rm G_F}\approx1.17\times 10^{-5}$ GeV$^{-2}$ is the Fermi constant.

 \begin{table}[t]
\begin{tabular}{c|c|c|c} \hline
Process & $(i,j)$ & Experimental bounds & References \\ \hline
$\mu^{-} \to e^{-} \gamma$ & $(\mu,e)$ &
	${\rm BR}(\mu \to e\gamma) < 4.2 \times 10^{-13}$  ($90\%$ CL)& \cite{MEG:2016leq} \\
$\tau^{-} \to e^{-} \gamma$ & $(\tau,e)$ &
	${\rm BR}(\tau \to e\gamma) < 3.3 \times 10^{-8}$  ($90\%$ CL)& \cite{BaBar:2009hkt} \\
$\tau^{-} \to \mu^{-} \gamma$ & $(\tau,\mu)$ &
	${\rm BR}(\tau \to \mu\gamma) < 4.4 \times 10^{-8}$ ($90\%$ CL) & \cite{BaBar:2009hkt}   \\ \hline
$\Delta a_e$ & $(e,e)$ &
$ (3.41\pm1.64) \times 10^{-13}$   ($1\sigma$) & \cite{Fan:2022eto}   \\ \hline%
$\Delta a_\mu$ & $(\mu,\mu)$ &
	$ (39\pm64) \times 10^{-11}$ ($1\sigma$) & \cite{Aliberti:2025beg}   \\ \hline
\end{tabular}
\caption{Summary for the experimental bounds of the LFV processes 
$\ell_\alpha \to \ell_\beta \gamma$ and lepton $g-2$.}
\label{tab:Cif}
\end{table}

The contributions to electron and muon $g-2$ are given by
\begin{align}
\Delta a_{ii}
\approx 
-
\frac{m_i^2}{8\pi^2} \sum_{\alpha} (y^\eta_{i\alpha}(y^\eta)^\dag_{\alpha i})G(M_{X_\alpha},m_{\eta^+}).
\label{damu}
\end{align}

\subsection{Dark matter candidate}
\label{sec:III}
Dark matter candidates in our model are $X_R$, $X'_R$, $\eta^0$, and $S_0$, whose stability is guaranteed by the $\mathbb{Z}_3$ symmetry. 
Here, we regard the lightest $X_R$ as the DM candidate, denoted by $\chi_R$ with mass $m_\chi$. This case is of particular interest since it only interacts via $y^\eta$ and thus the dark matter relic abundance is related to the neutrino mass generation mechanism.
The cross section for the relic density in the thermal freezeout scenario is given by
\begin{align}
(\sigma v_{\rm rel})&\approx \frac{m^2_\chi}{48\pi}\sum_{i,j=1}^3
\left[
\frac{m^4_\chi+m_{\eta_0}^4}{(m^2_\chi+m_{\eta_0}^2)^4}
\left|\sum_{a,b=1}^3 U^\dag_{ia} y^\eta_{a1}(y^\eta)^\dag_{1b}U_{bj} \right|^2
+
\frac{m^4_\chi+m_{\eta^+}^4}{(m^2_\chi+m_{\eta^+}^2)^4}
\left|y^\eta_{i1}(y^\eta)^\dag_{1j}\right|^2
\right] v_{\rm rel}^2\nn\\
&\equiv
\frac{m^2_\chi}{48\pi}
|y^{\eta\dagger}y^\eta|_{11}^2
\left[
\frac{m^4_\chi+m_{\eta_0}^4}{(m^2_\chi+m_{\eta_0}^2)^4}
+
\frac{m^4_\chi+m_{\eta^+}^4}{(m^2_\chi+m_{\eta^+}^2)^4}
\right] v_{\rm rel}^2,
\end{align}
where we expand the cross section in terms of relative velocity $v_{\rm rel}^2\approx0.2$~\cite{Cannoni:2013bza}, {and have used the following relation: $\sum_{ij}|(U^\dagger y^\eta)_{i1}(U^\dagger y^\eta)_{j1}|=\sum_{ij}\left|y^\eta_{i1}(y^\eta)^\dag_{1j}\right|^2\equiv |y^{\eta\dagger}y^\eta|_{11}^2$.}
To satisfy the correct relic density at $2\sigma$~\cite{Planck:2018vyg}; $0.118\lesssim\Omega h^2\lesssim 0.122$,
the cross section is within the range of
\begin{align}
1.71\times10^{-9}\ [{\rm GeV}^{-2}] \lesssim
(\sigma v_{\rm rel})\lesssim 2.03\times10^{-9}\ [{\rm GeV}^{-2}].
\end{align}

 \section{Numerical analyses}
 \label{sect.3}
 
In this section, we perform numerical analyses considering all the constraints that we have discussed above, and show the allowed region. Specifically, we consider neutrino oscillation data, non-unitarily bound, the sum of the neutrino masses, lepton flavor violations, lepton anomalous magnetic dipole moment, dark matter with correct relic density assuming one of $X_R$ has the lightest mass among the DM candidates.

For the analysis, we randomly scan our input parameters within the following ranges:
\begin{align}
& |y_{a\alpha}|=[10^{-3},\sqrt{4\pi}],\quad |\tilde\theta_{12,13,23}|=[10^{-1},10],\quad
(\alpha,\beta)=[-\pi,\pi],\\
& M_{X_1}(\equiv m_\chi)=[10^2,10^4][{\rm GeV}],\quad 
(M_{X_2},m_{\eta},m_{S_0},M_{N_1})=[1.2\times m_\chi,10^5][{\rm GeV}],\\ 
&M_{X_3} =[1.2\times M_{X_2},10^5][{\rm GeV}],\quad 
M_{N_{2}} =[1.2\times M_{N_1},10^5][{\rm GeV}],\\
&M_{N_{3}} =[1.2\times M_{N_2},10^5][{\rm GeV}],\quad \mu=[10^{-3},10^{3}][{\rm GeV}],\\
&D_{\nu_{1(3)}}=[0.1,100][{\rm meV}]\ {\rm for}\ {\rm NH(IH)},
\end{align}
where we take $m_{\eta}\equiv m_{\eta^+}=m_{\eta_0}$ to conservatively evade the constraint of the oblique parameters, and $\alpha,\beta$ are Majorana phases defined by diag$[1,e^{i\alpha/2},e^{i\beta/2}]$.
Here, $\tilde\theta_{12,13,23}$ are complex mixing angles of the orthogonal matrix ${\cal O}$ in Eq.(\ref{eq:ci}) using the same parametrization for $U_{\rm PMNS}$.
Note that once we choose the lightest neutrino mass, all the neutrino mass eigenvalues are fixed by the experimental values of neutrino mass square differences. We will use the best fit values of neutrino oscillation data in Nufit6.0~\cite{Esteban:2024eli} for NH and IH.

\subsection{NH}

\begin{figure}[tb]\begin{center}
\includegraphics[width=80mm]{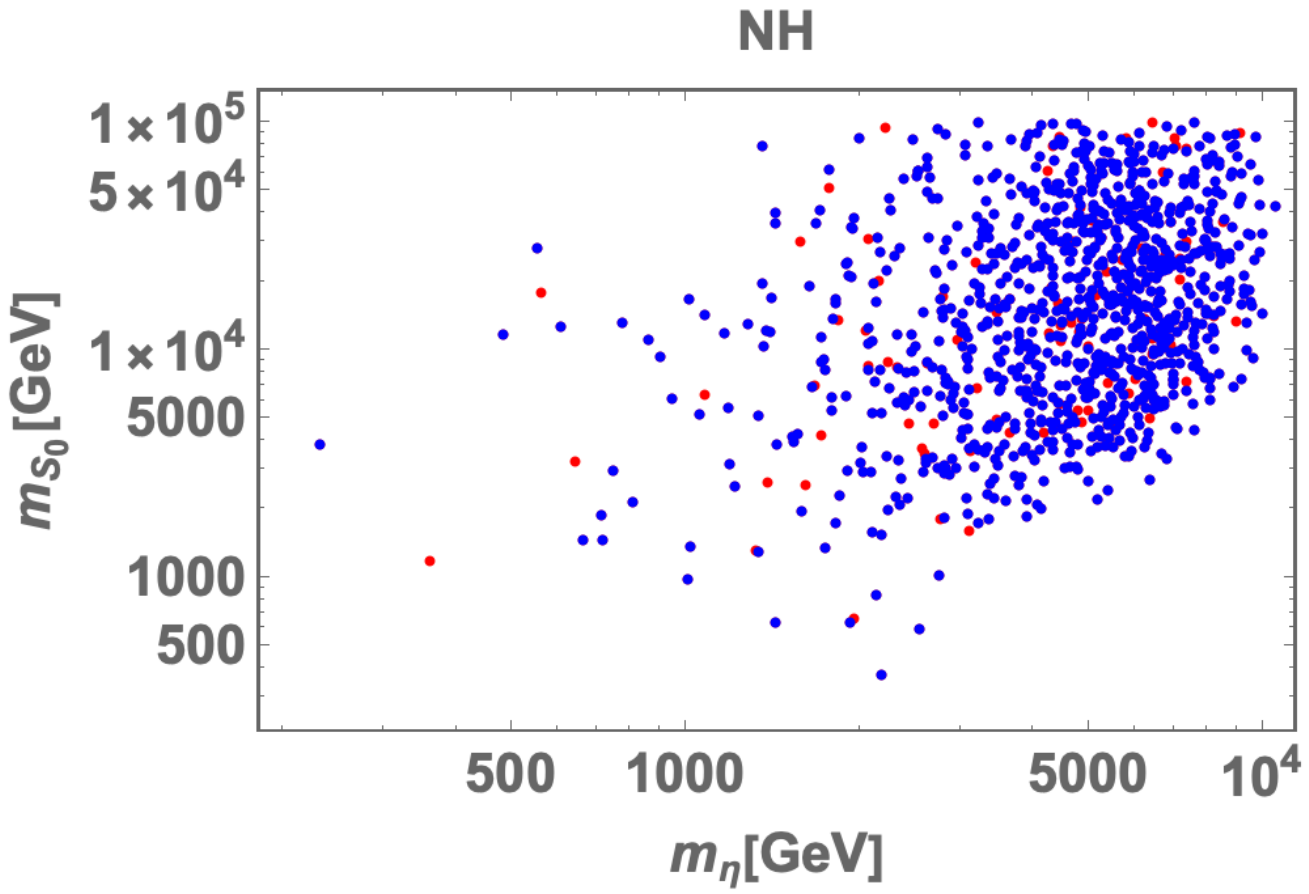}
\caption{Allowed regions of $m_\eta$ and $m_{S_0}$ to satisfy all the conditions which we discussed above, where red plots do not satisfy $\sum_i D_i\le120$ meV. } 
\label{fig:nh1}
\end{center}\end{figure}
In Fig.~\ref{fig:nh1}, we show allowed region of $m_\eta$ and $m_{S_0}$. 
Here and hereafter, we plot the points in blue when the condition $\sum_i D_i\le120$ meV is satisfied, and in red otherwise (we plot only $\sum_i D_i\leq200$ meV). We see that
$m_{S_0}$ scatters over the whole region, while $m_\eta$ seems to have an upper bound of $10^4$ GeV.

\begin{figure}[tb]\begin{center}
\includegraphics[width=53mm]{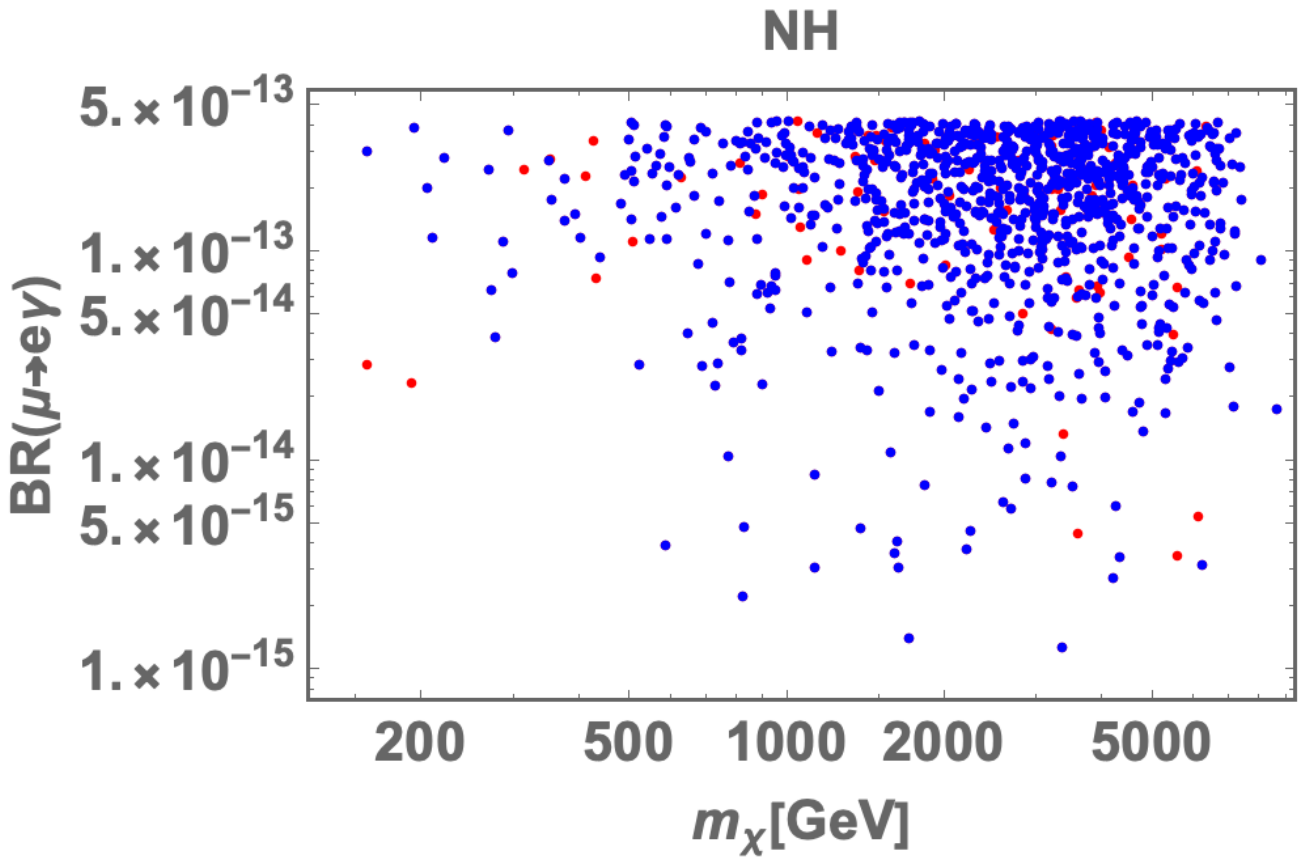}
\includegraphics[width=53mm]{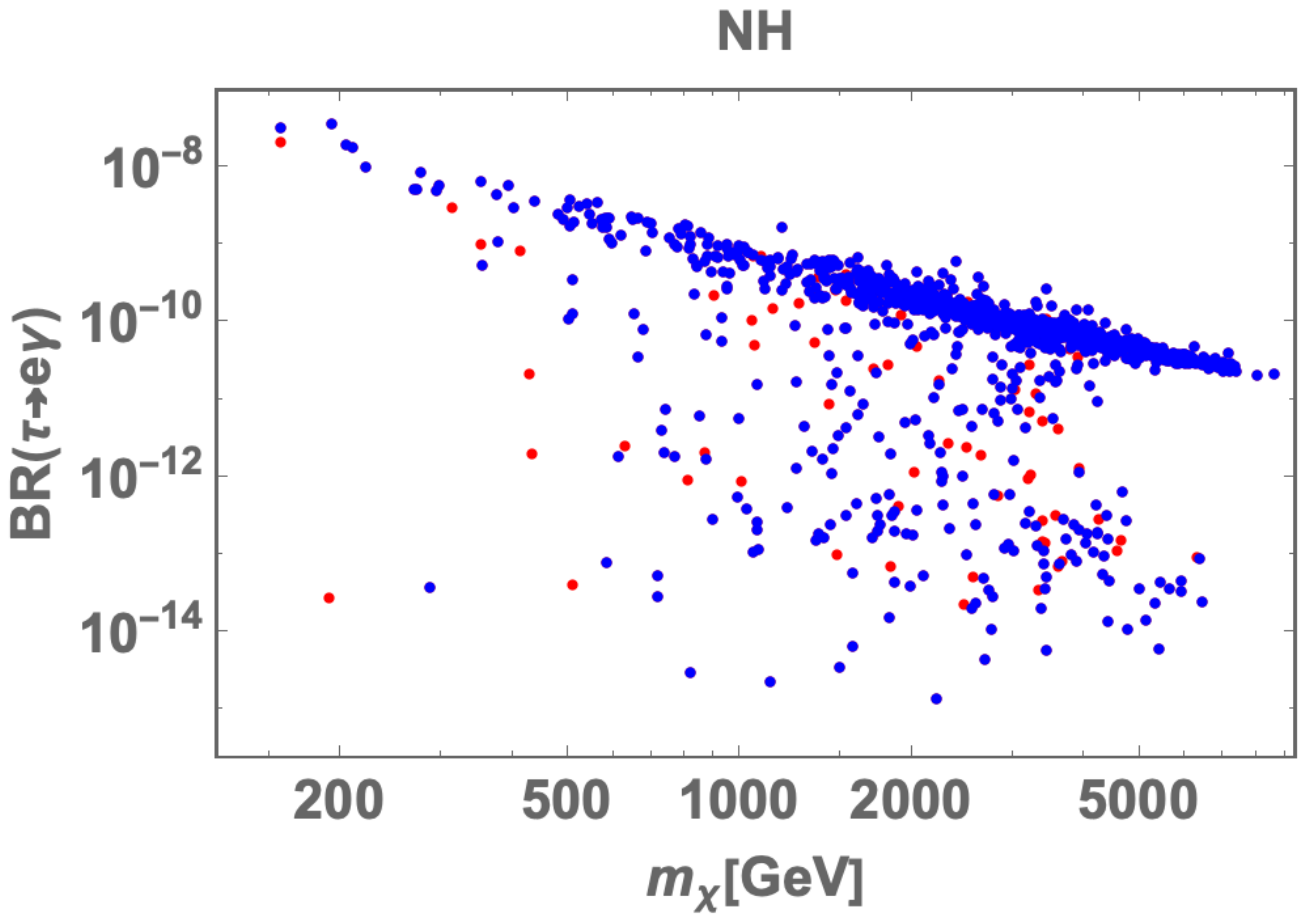}
\includegraphics[width=53mm]{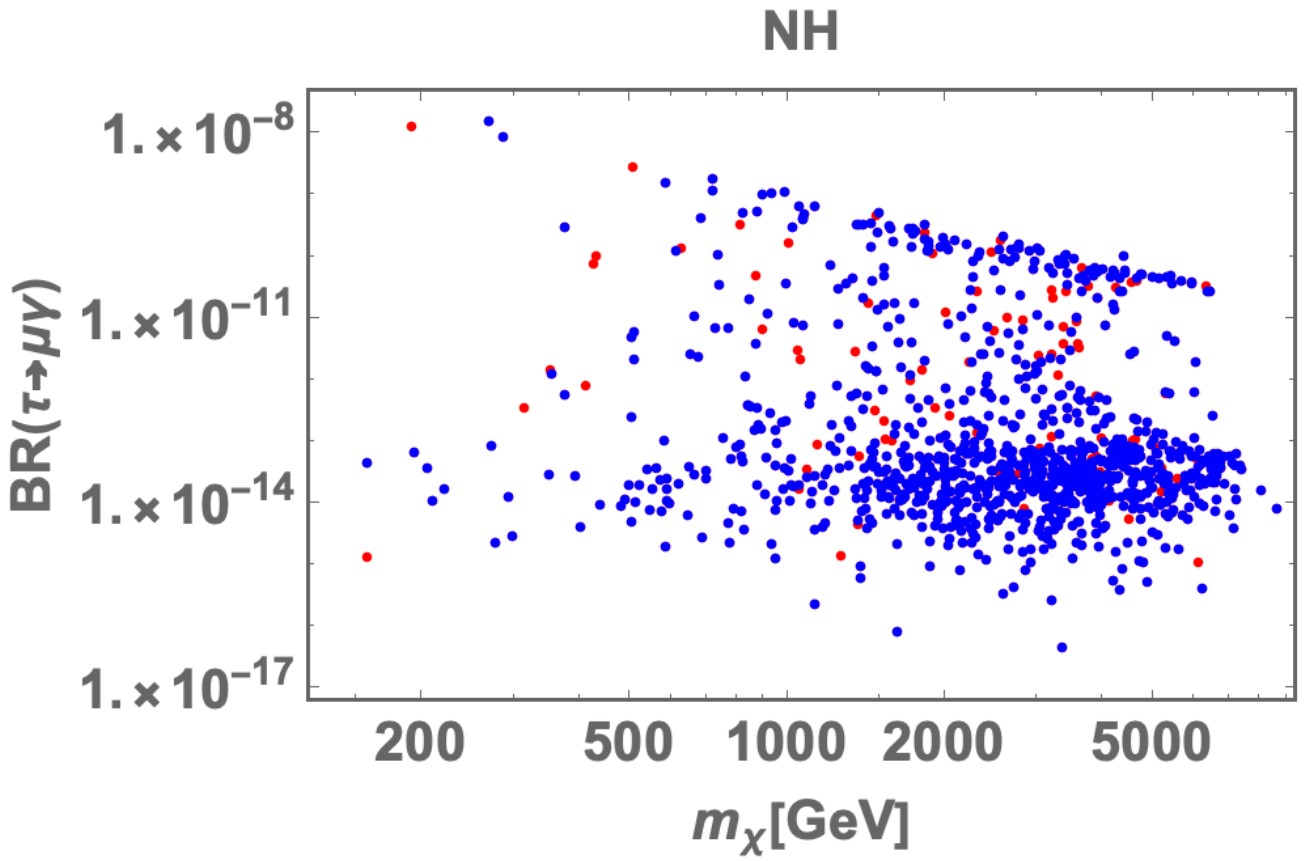}
\caption{Values of LFVs for a given DM mass. The color legend is the same as the one of Fig.~\ref{fig:nh1}.} 
\label{fig:nh_lfvs_nh}
\end{center}\end{figure}
In Fig.~\ref{fig:nh_lfvs_nh}, we show values of LFVs for a given DM mass. 
The figures suggest that a lot of allowed points are around the boundary of the experimental limit for ${\rm BR}(\mu\to e\gamma)$.
On the other hand, ${\rm BR}(\tau\to \mu\gamma)\simeq10^{-14}$ is favored
due to its stronger correlation to ${\rm BR}(\mu\to e\gamma)$.
The lower DM mass is not favored by LFVs, excluding a large number of model points for $m_\chi\lesssim500$ GeV.

\begin{figure}[tb]\begin{center}
\includegraphics[width=53mm]{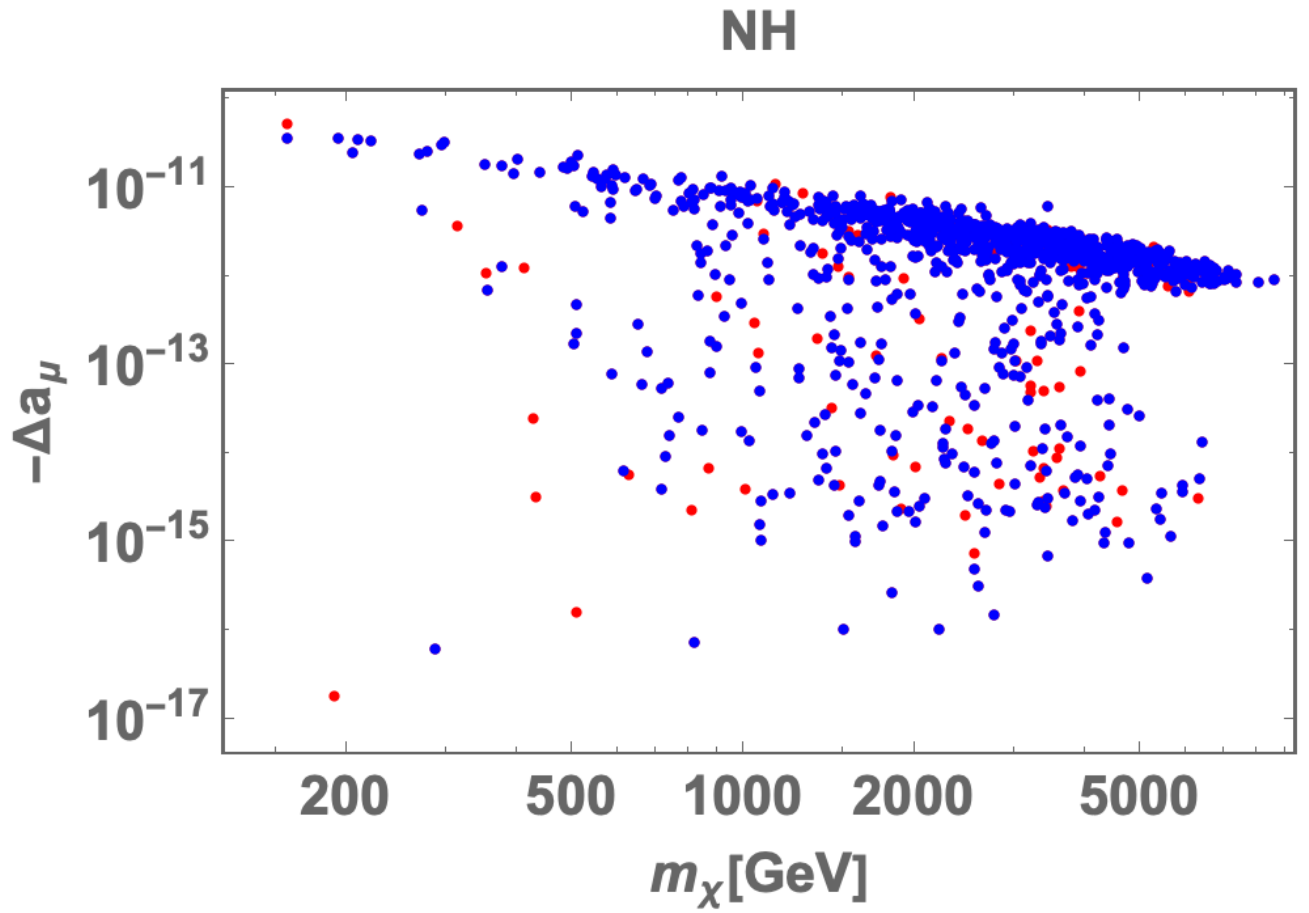}
\includegraphics[width=53mm]{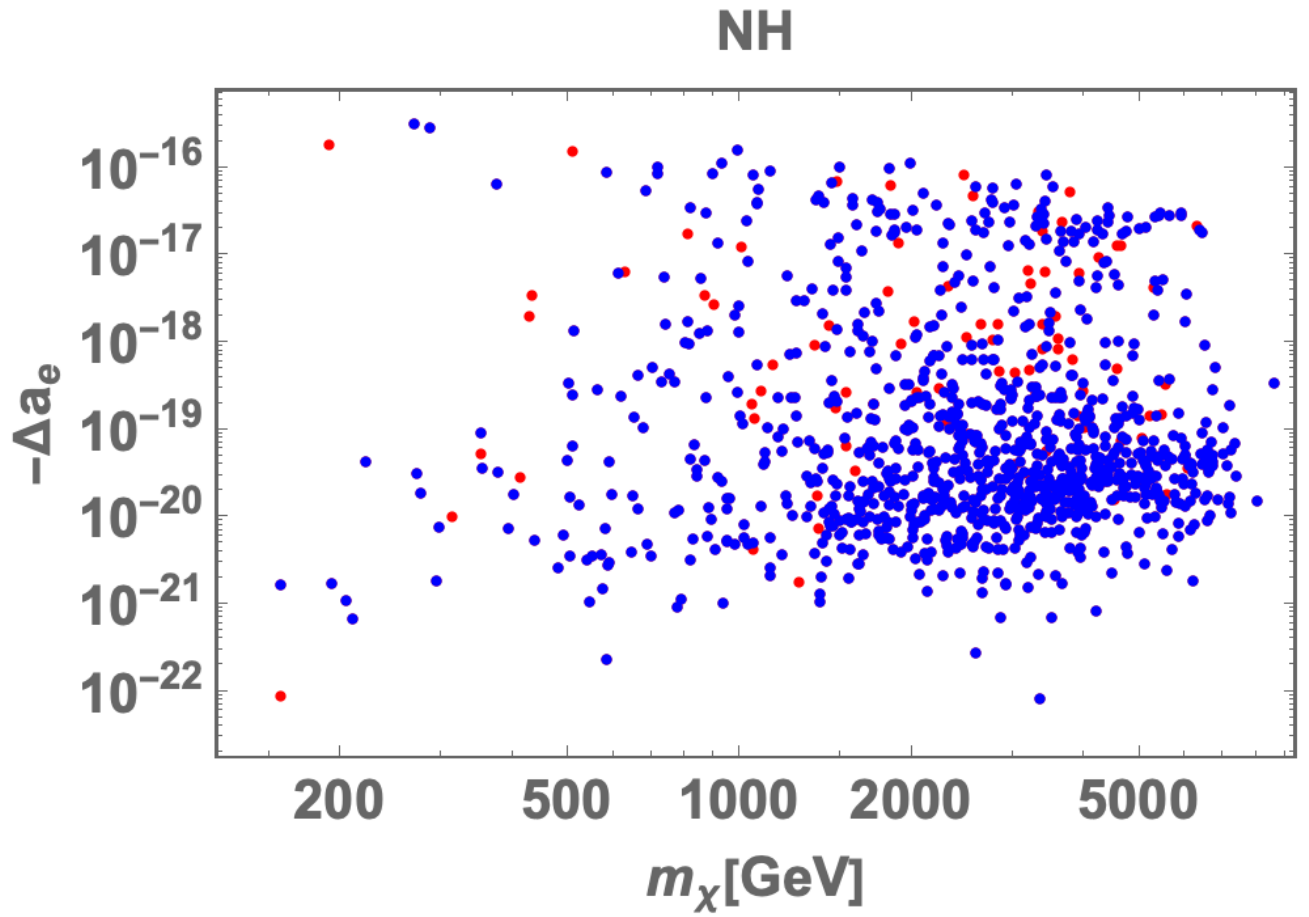}
\caption{Prediction of lepton $g-2$ for a given DM mass. The color legend is the same as the one of Fig.~\ref{fig:nh1}. } 
\label{fig:daell_nh}
\end{center}\end{figure}
In Fig.~\ref{fig:daell_nh}, we show prediction of lepton $g-2$ for a given DM mass.
The contributions to both electron and muon $g-2$ are totally within the current experimental bounds.

\begin{figure}[tb]\begin{center}
\includegraphics[width=53mm]{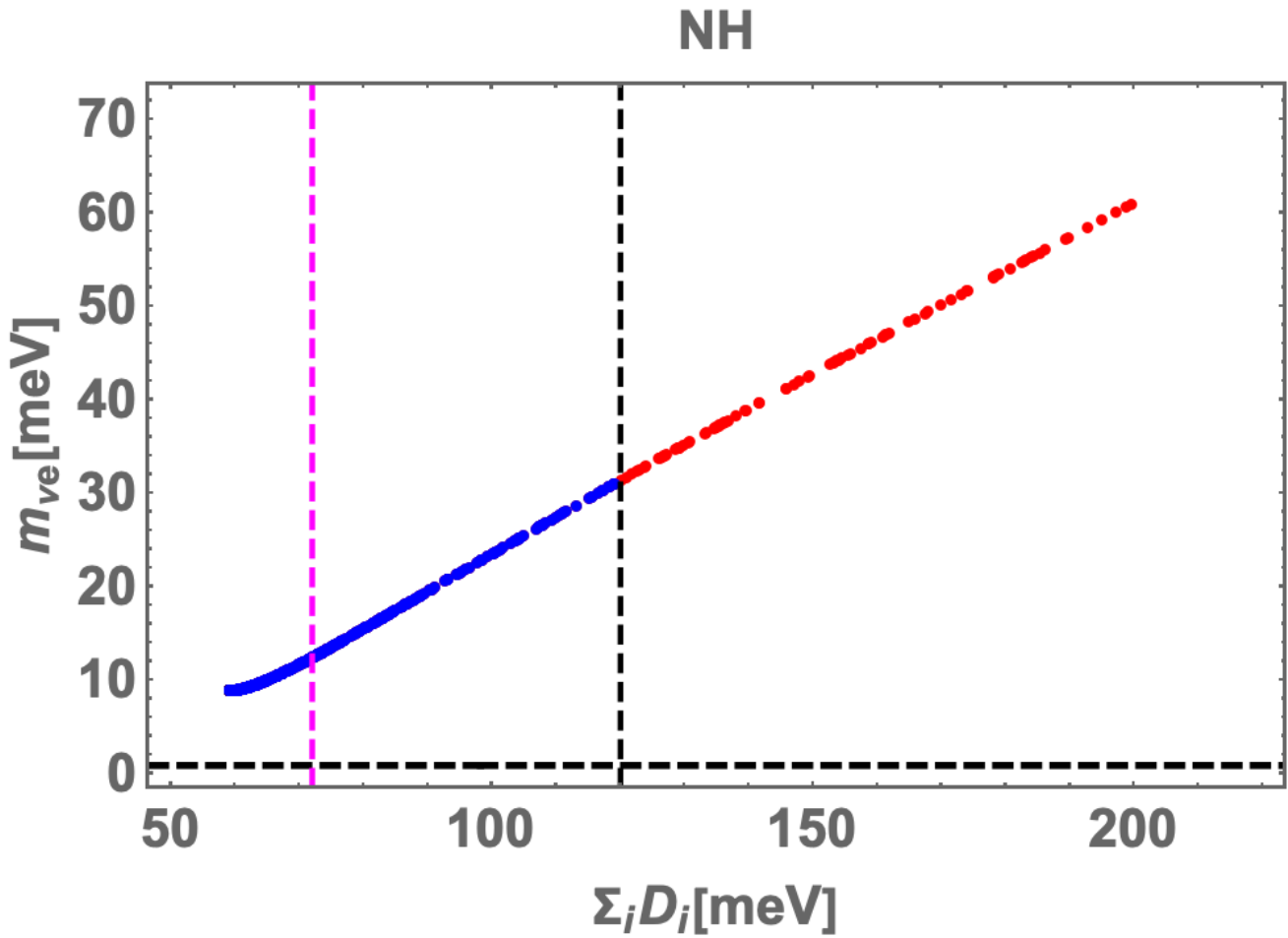}
\includegraphics[width=53mm]{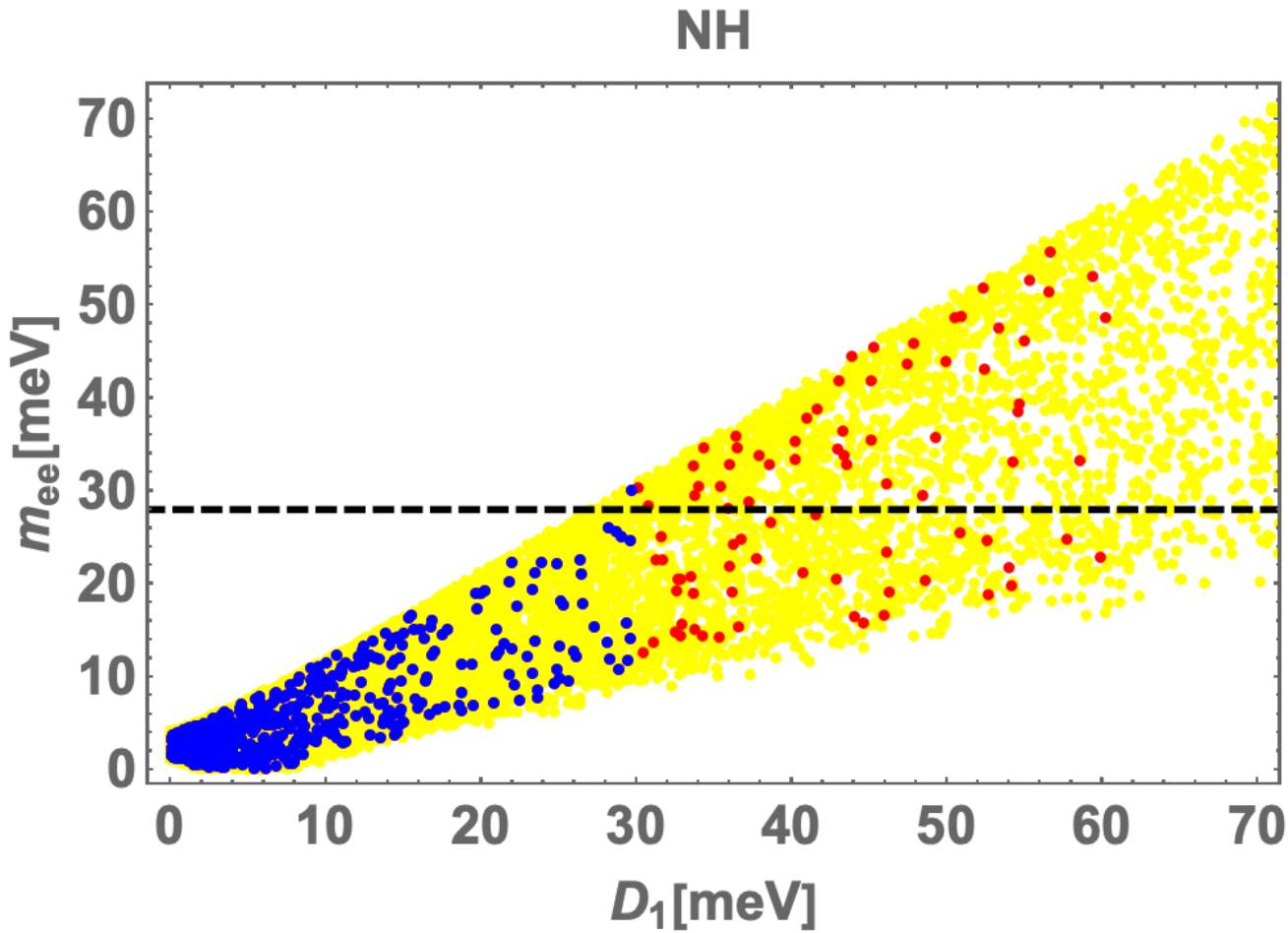}
\includegraphics[width=53mm]{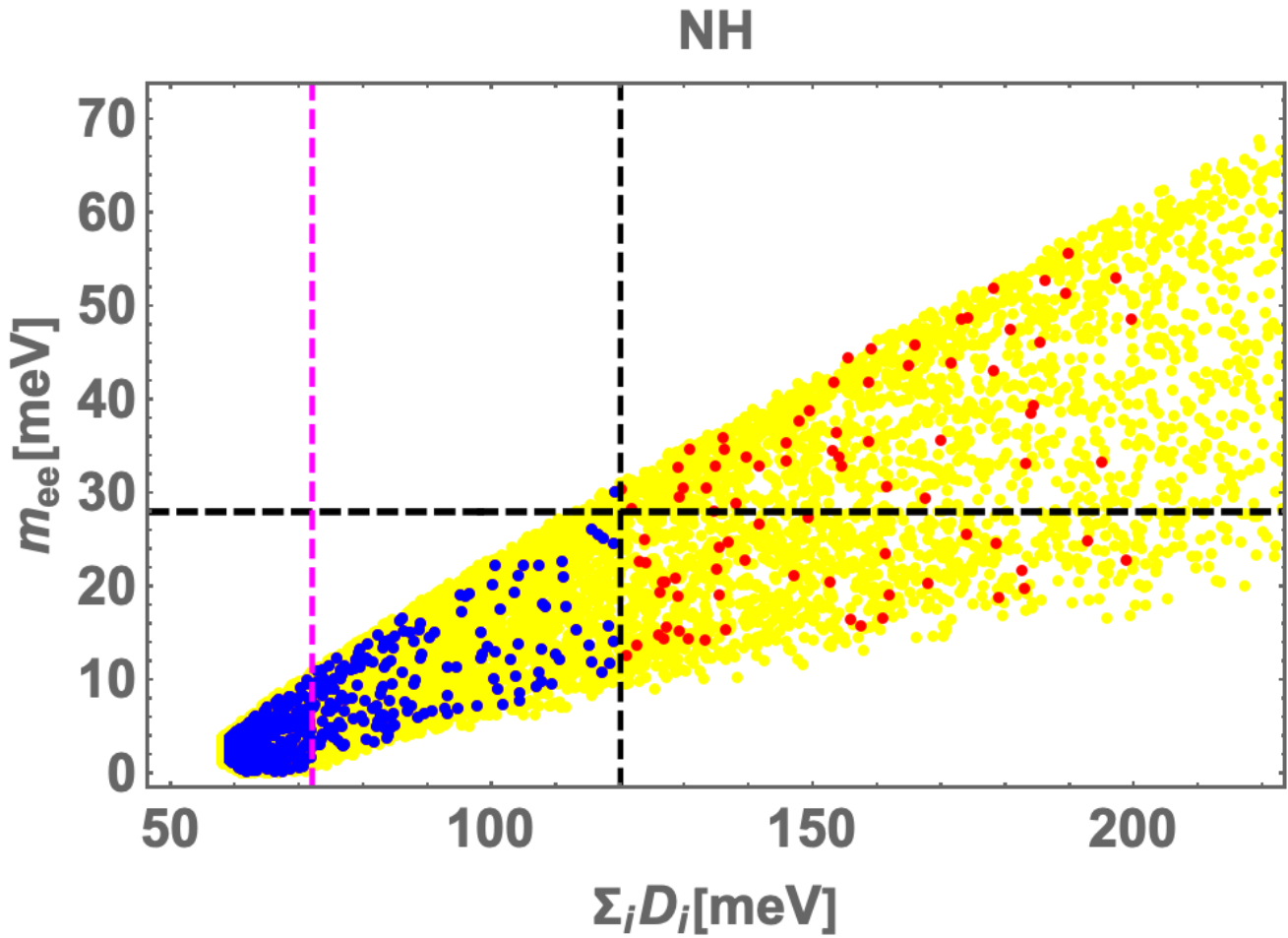}
\caption{Predictions related to active neutrino masses where the color legend is the same as the one of Fig.~\ref{fig:nh1}. The yellow region represents the parameter space consistent with the Nufit 6.0 global fit.} 
\label{fig:neut_nh}
\end{center}\end{figure}
In Fig.~\ref{fig:neut_nh}, we show predictions related to active neutrino masses.
The left panel shows the scatter plot in the $m_{\nu e}$ and $\sum_i D_i$ plane.
The vertical magenta dashed line represents upper bound on the combination of DESI  and CMB, $\sum D_i\le72$ meV, while the vertical black one shows $\sum_i D_i\le120$ meV. The horizontal dashed black one is lower bound on $m_{\nu e}$ (0.85 meV). Since $m_{\nu e}$ does not depend on phases, the scatter points are distributed along a narrow line.
The middle panel is the scatter plot in the $m_{ee}$ and the lightest active neutrino mass $D_1$ plane. The horizontal dashed black line is minimum upper bound on $m_{ee}$ (28 meV). The yellow region represents the parameter space consistent with the Nufit 6.0 global fit, excluding other constraints.
This figure suggests that almost all the blue plots; $\sum_i D_i\le120$ meV, are lower than $m_{ee}=28$ meV.
The right panel is the scatter plot in the $m_{ee}$ and $\sum_i D_i$ plane. All the lines and yellow regions are the same as in the middle panel.
This figure implies that a lot of points satisfy $\sum_i D_i\le72$ meV.

\subsection{IH}
\begin{figure}[tb]\begin{center}
\includegraphics[width=80mm]{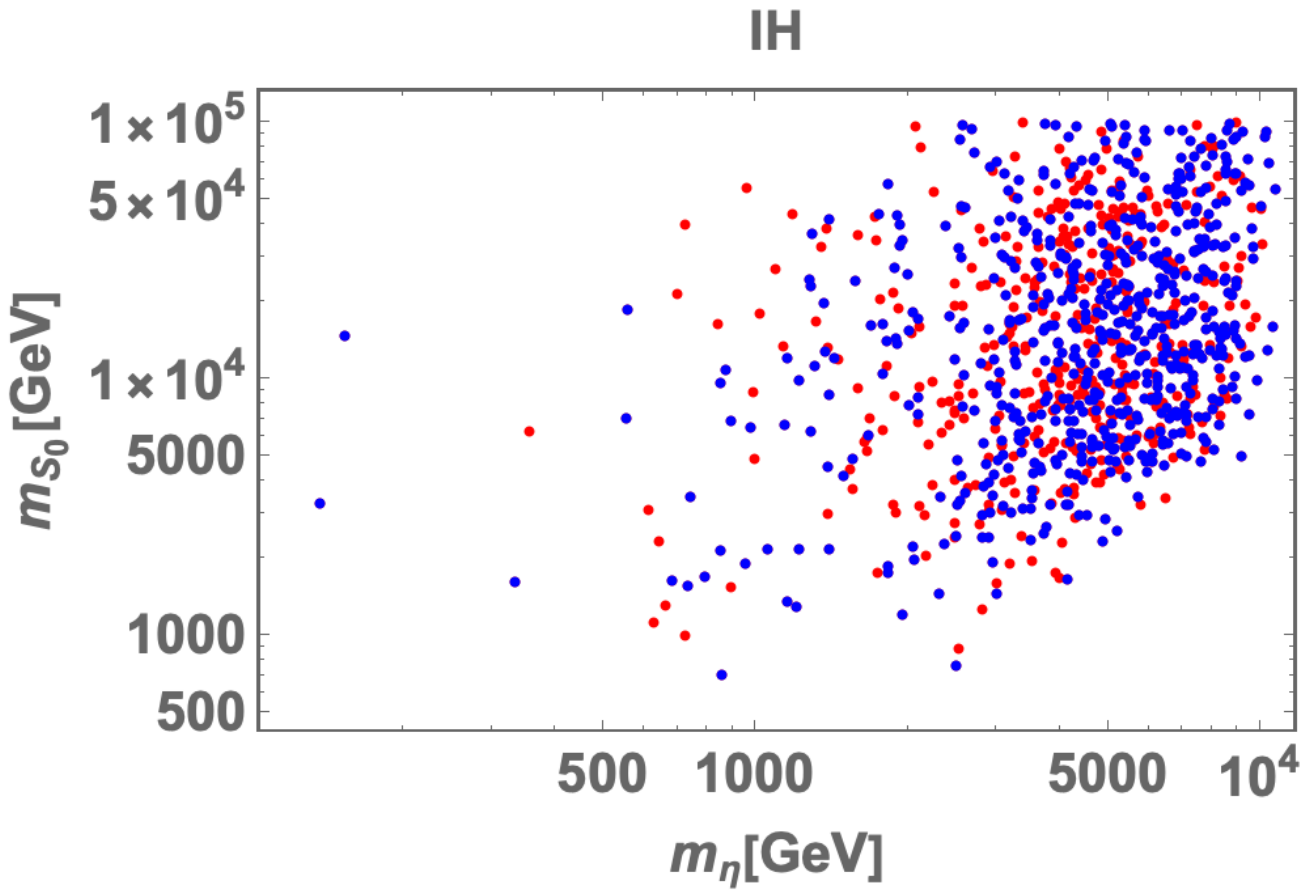}
\caption{The same figure as Fig.~\ref{fig:nh1} but for the inverted hierarchy.} 
\label{fig:ih1}
\end{center}\end{figure}

\begin{figure}[tb]\begin{center}
\includegraphics[width=53mm]{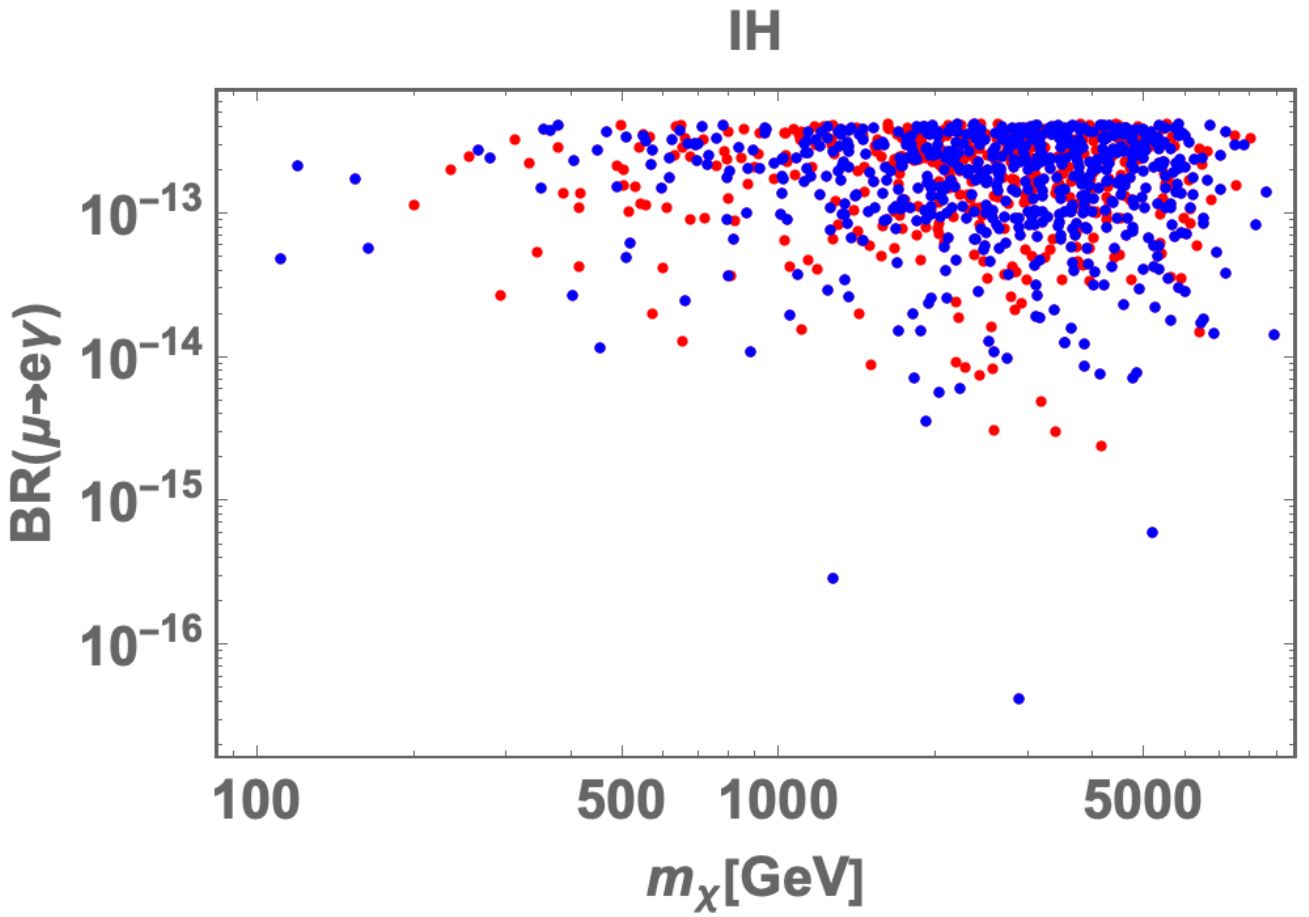}
\includegraphics[width=53mm]{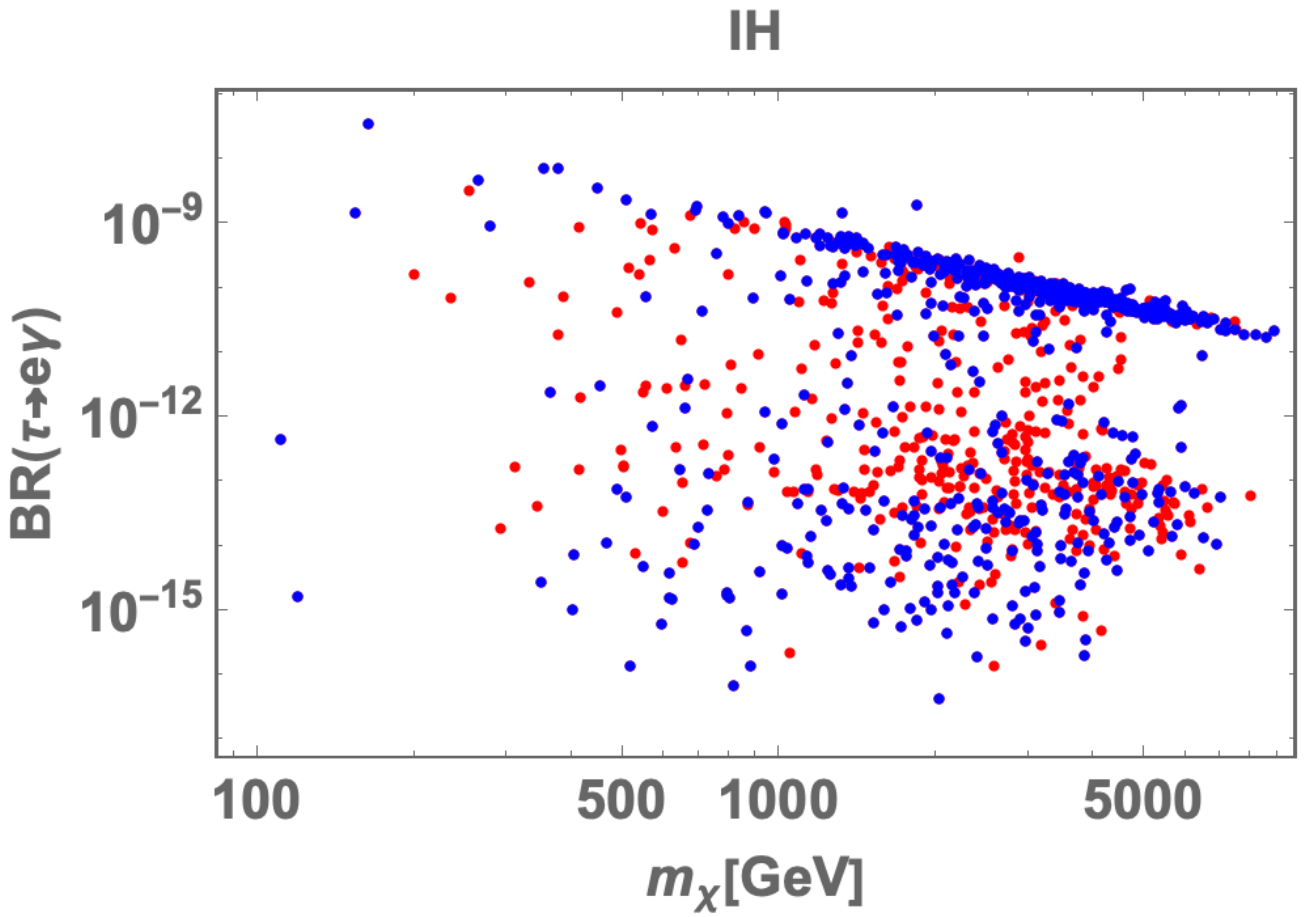}
\includegraphics[width=53mm]{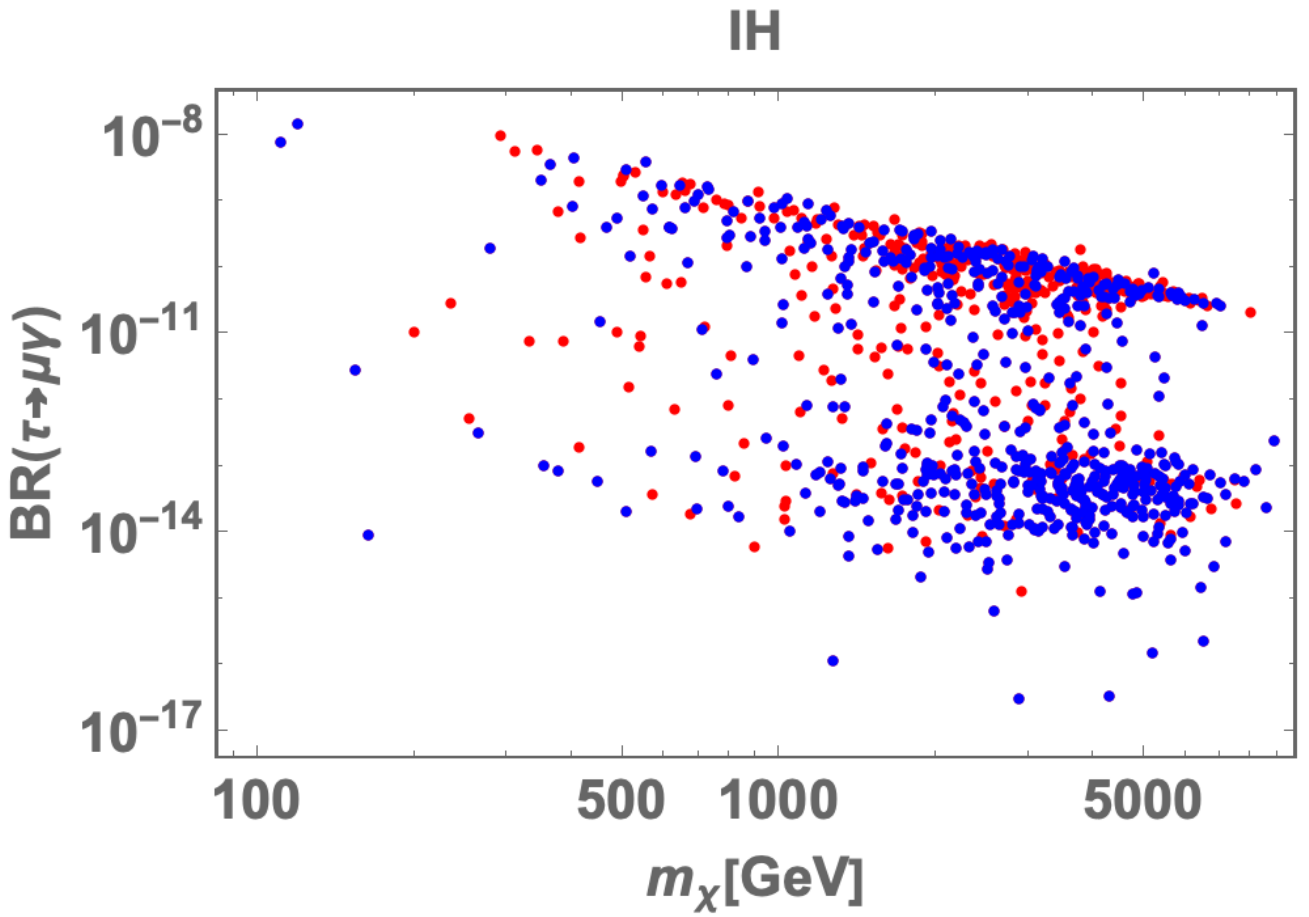}
\caption{The same figure as Fig.~\ref{fig:nh_lfvs_nh} but for the inverted hierarchy.} 
\label{fig:nh_lfvs_ih}
\end{center}\end{figure}

\begin{figure}[tb]\begin{center}
\includegraphics[width=53mm]{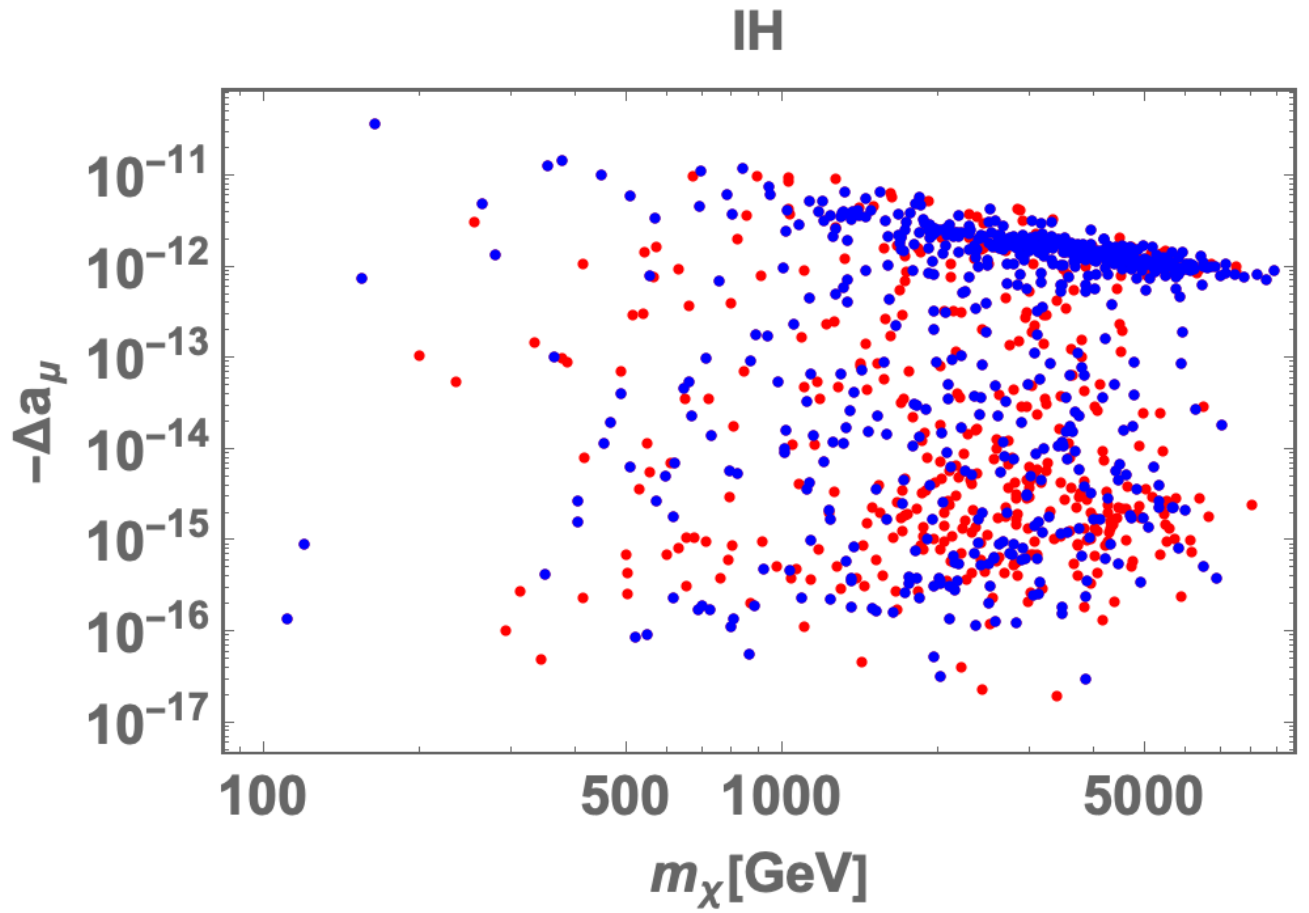}
\includegraphics[width=53mm]{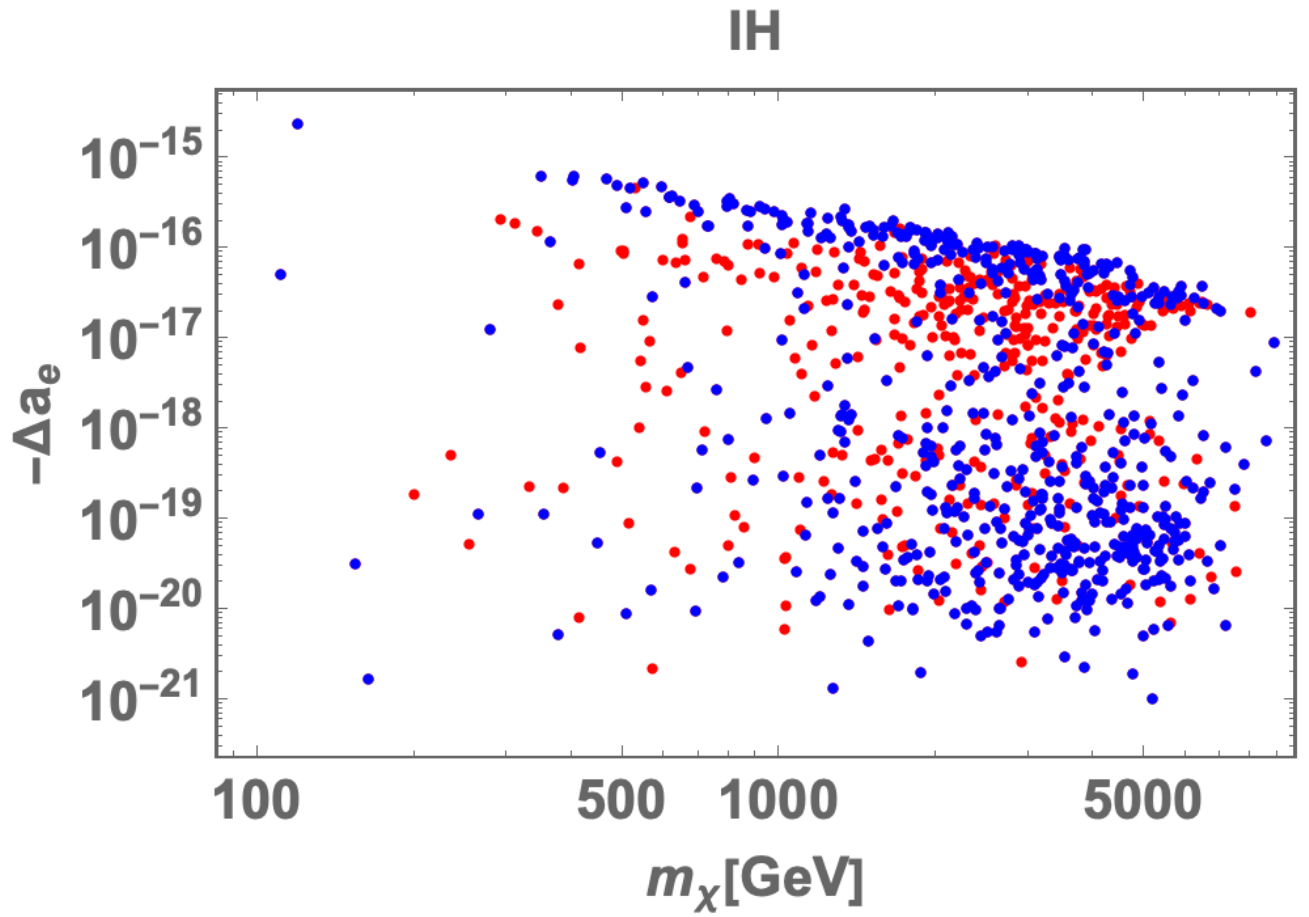}
\caption{The same figure as Fig.~\ref{fig:daell_nh} but for the inverted hierarchy.} 
\label{fig:daell_ih}
\end{center}\end{figure}

\begin{figure}[tb]\begin{center}
\includegraphics[width=53mm]{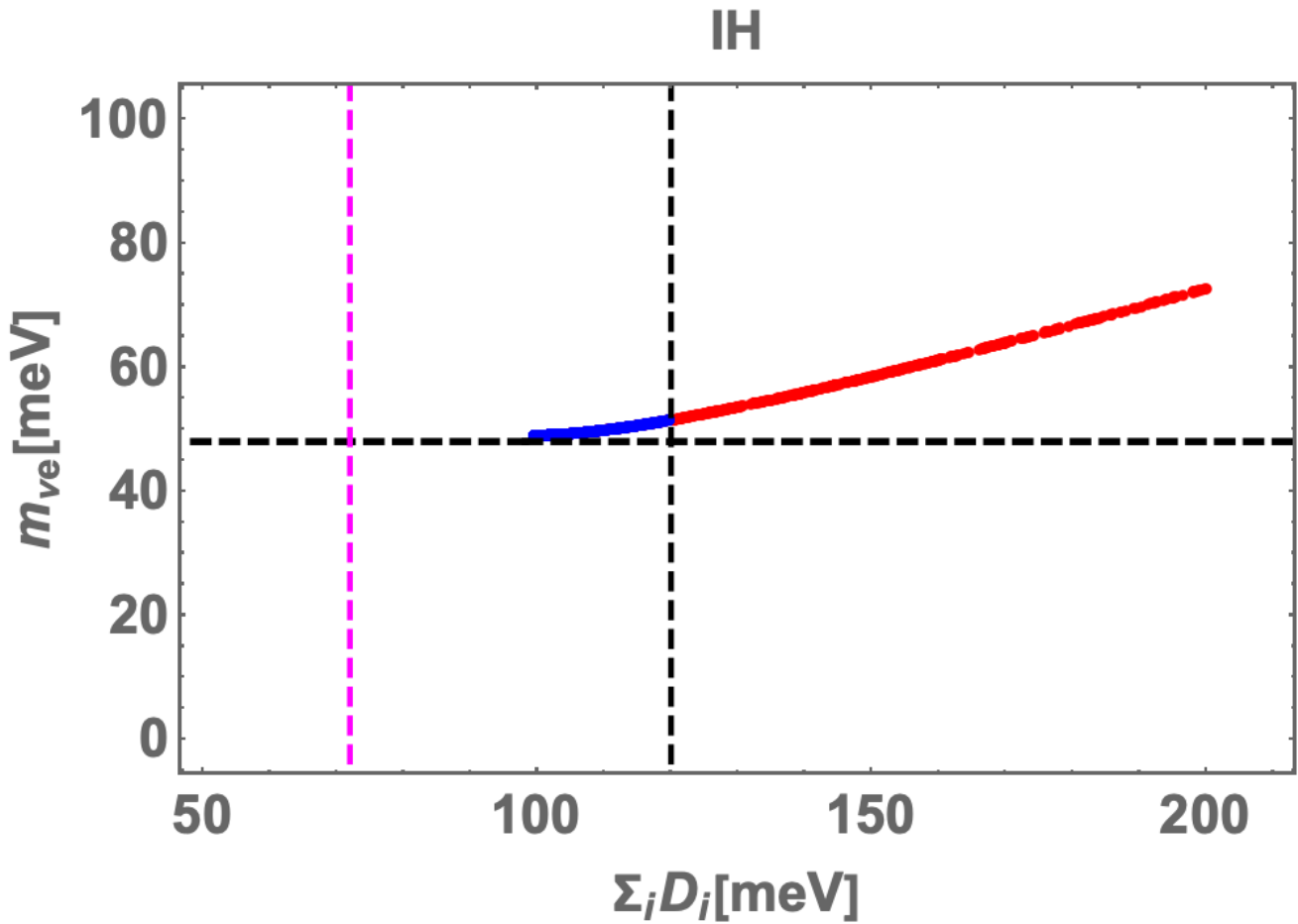}
\includegraphics[width=53mm]{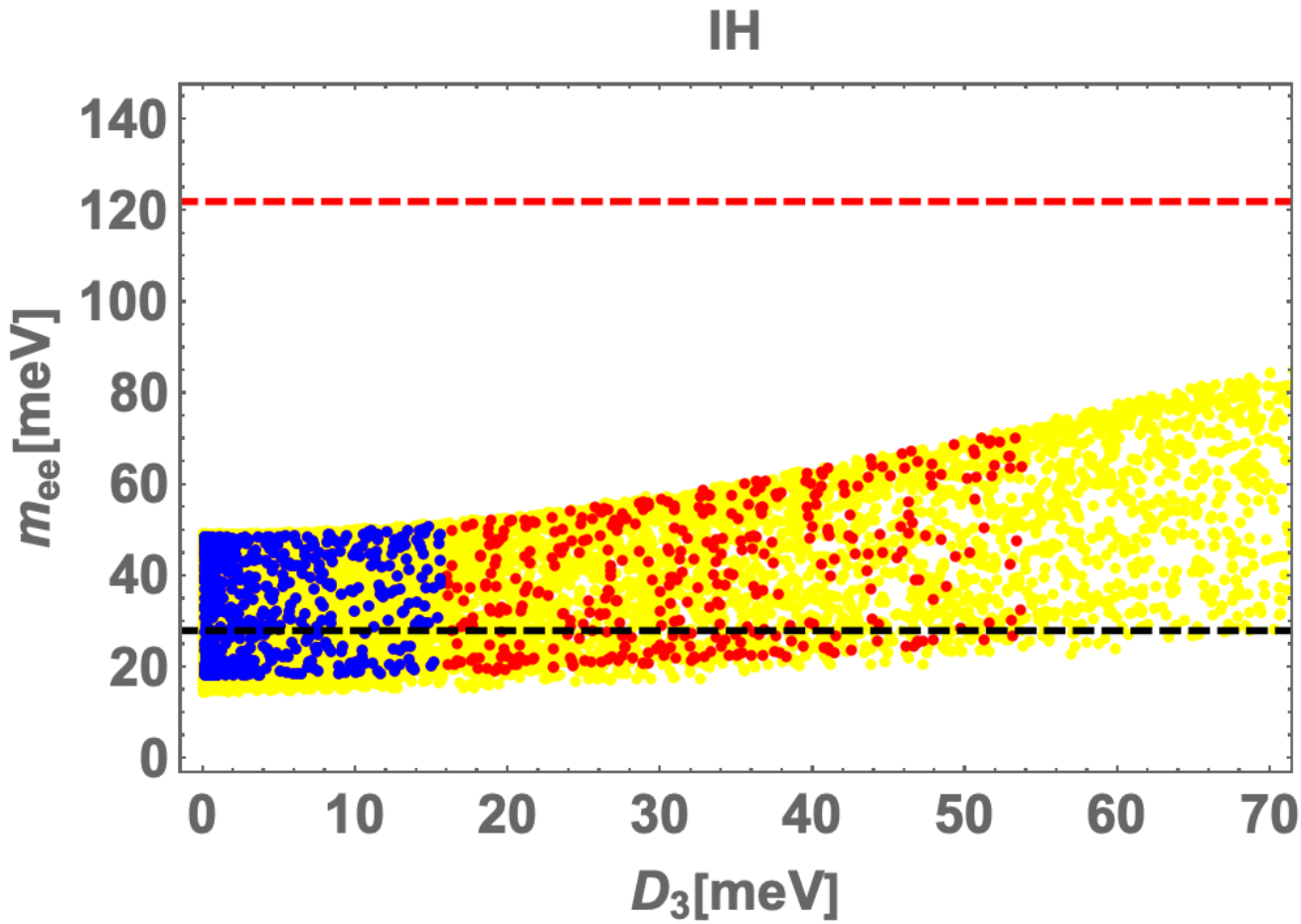}
\includegraphics[width=53mm]{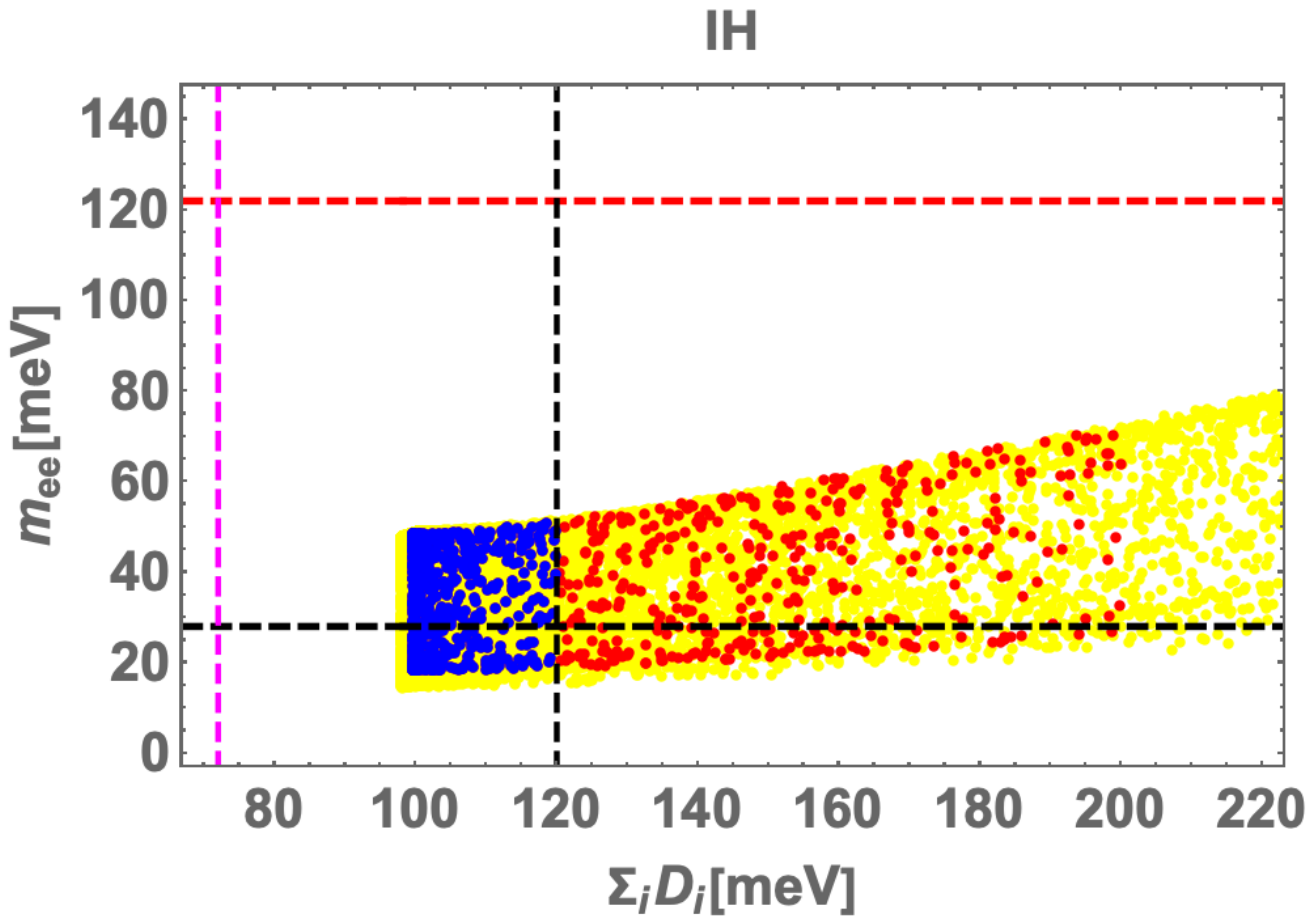}
\caption{The same figure as Fig.~\ref{fig:neut_nh} but for the inverted hierarchy.} 
\label{fig:neut_ih}
\end{center}\end{figure}

Figs.~\ref{fig:ih1}, \ref{fig:nh_lfvs_ih}, \ref{fig:daell_ih}, and \ref{fig:neut_ih} show the corresponding results for the IH case, corresponding to Figs.~\ref{fig:nh1}, \ref{fig:nh_lfvs_nh}, \ref{fig:daell_nh}, and \ref{fig:neut_nh}, respectively. The trends observed in LFV and lepton $g-2$ that we have discussed above also exist in the IH case.
However, we observed a slight shift in the preferred region in the neutrino sector: the lightest neutrino mass shifts to smaller values, $D_3\lesssim16 (54)$ meV for $\sum_i\lesssim 120(200)$ meV; the lower bound on the sum of the neutrino mass becomes stronger, $100\ {\rm meV}\lesssim \sum_i D_i$; and $m_{ee}$ develops a lower bound, $18\ {\rm meV}\lesssim m_{ee}$, with a weaker dependence on $D_3$ and $\sum_iD_i$.


\section{Summary and discussion}
We have proposed a novel realization of the linear seesaw mechanism by employing a $\mathbb Z_3$ gauging Tambara-Yamagami fusion rule (TY) with the assistance of $\mathbb Z_3$ symmetry.
In this framework, mass hierarchies of Eq.~\eqref{eq:neutmass-order_exp} are theoretically explained via a loop suppression. It is noteworthy that the TY symmetry is dynamically broken to generate the hierarchical structure, while it controls the interaction terms to realize the structure of the linear seesaw.
Due to this structure, the active neutrino masses are generated at the two-loop level, which naturally predicts new particles around the TeV scale.
In addition to the mass and mixing of the active neutrinos, we have also taken into consideration of lepton flavor violations, lepton $g-2$, and dark matter.
In particular, we are interested in the case where the lightest Majorana fermion explains the dark matter relic abundance.
In our numerical analysis, we have searched for allowed regions satisfying all the constraints; neutrino oscillation data, non-unitarity bound, three lepton flavor violations,
muon/electron $g-2$, observed relic density of dark matter.
We have demonstrated that the model has a large parameter space and found some tendencies of the observables for NH and IH.

\begin{acknowledgments}
 HO is supported by Zhongyuan Talent (Talent Recruitment Series) Foreign Experts Project. YS is supported by the Slovenian Research Agency under the research grant J1-4389.
\end{acknowledgments}

\appendix

\section{Fusion rule of ${\mathbb Z_3}$ gauging of Tambara-Yamagami fusuion rule }
\label{app}
Here, we show multiplication rules of $\mathbb Z_3$ gauging Tambara-Yamagami fusuion rule (TY) where TY consists of four commutable generators $\{{ \mathbbm{1},a,a^*,n}\}$~\cite{Dong:2025jra}:
\begin{align}
\begin{array}{lll}
n \otimes n = \mathbbm{1} \oplus a \oplus a^* \, , ~~~~~ & a \otimes a^* = \mathbbm{1} \, , ~~~~~ & a \otimes a =a^* \, , \\[0.3ex]
a^* \otimes a^* = a \, , ~~~~~ & a^* \otimes n = a^* \, , ~~~~~ & a \otimes n = a \, .
\end{array}
\label{eq:fusionrule}
\end{align} 
Note here that $n$ is a self-conjugate generator that does not have its inverse generator.

\bibliography{ctma4.bib}

\end{document}